# Resonant photonuclear isotope detection using medium-energy photon beam

Hiroyasu Ejiri[1,2,*] and Tatsushi Shima[1]

[1]*Research Center for Nuclear Physics, Osaka University, Ibaraki, Osaka 567-0047, Japan*
[2]*Nuclear Science, Czech Technical University, Prague, Czech Republic*

Resonant photonuclear isotope detection (RPID) is a nondestructive detection/assay of nuclear isotopes by measuring $\gamma$ rays following photonuclear reaction products. Medium-energy wideband photons of $E_\gamma = 12$–16 MeV are used for the photonuclear $(\gamma, n)$ reactions and $\gamma$ rays characteristic of the reaction products are measured by means of high-sensitivity Ge detectors. Impurities of stable and radioactive isotopes of the orders of $\mu$gr—ngr and ppm—ppb are investigated. RPID is used to study nuclear isotopes of astronuclear and particle physics interests and those of geological and historical interests. It is used to identify radioactive isotopes of fission products as well.

## I. INTRODUCTION

Nondestructive high-sensitivity detection of nuclear isotopes is of great interest for basic and applied science. This report aims to show that resonant photonuclear isotope detection (RPID) is very powerful for nondestructive studies of nuclear isotopes with sensitivities of the orders of ppm—ppb and $\mu$gr—ngr.

RPID uses resonant photonuclear isotope transmutations (RPIT) [1,2] to transmute impurity isotopes $X$ to radioactive isotopes (RIs) $X'$ by photonuclear reactions via E1 giant resonance and measure nuclear $\gamma$ rays from RIs by high-sensitivity Ge detectors. The large cross section of photonuclear reactions via E1 giant resonance and the high energy-resolution measurement of $\gamma$ rays characteristic of the photonuclear reaction products are key points of the present high-sensitivity detection/assay of nuclear isotopes. Here medium-energy photon beams for photonuclear reactions are obtained by intense laser photons backscattered off GeV electrons.

Neutron activation analysis has extensively been used for high-sensitivity isotope assay. It, however, is used only for isotopes with large neutron-capture cross sections. Since photonuclear reactions and neutron-capture reactions are quite different, they are complementary to each other. Accelerated mass spectroscopy is also used for high-sensitivity detection of isotopes in case of destructive assay.

RPID is very interesting for studying rare isotopes associated with nuclear processes of astronuclear and particle physics interests, small components of nuclear isotopes of geological and historical interests, small nuclear and atomic impurities of material science interests, and for other basic and applied science and technology.

## II. UNIQUE FEATURES OF RPID

RPID via the E1 giant resonance has such unique features that make it realistic to study nuclear isotopes/impurities of ppm—ppb level by measuring delayed $\gamma$ rays from RIs produced by the photonuclear reactions. They are as follows.

(i) The photonuclear reaction cross section via the E1 giant resonance is as large as $\sigma \approx 0.3A$ fm$^2$, with $A$ being the mass number, since many nucleons are resonantly excited. Thus, it can be used to study/assay all nuclear isotopes if $\gamma$ rays from the reaction products are well measured.

(ii) Since the resonance width is as broad as $\Gamma \approx 5$ MeV, a wideband photon beam with a bandwidth of $\Delta E \approx 5$ MeV is used. Medium-energy photons are produced by laser photons scattered off GeV electrons. Then about half of the photons are in the energy window, and are used to excite the E1 giant resonance.

(iii) Photonuclear reactions used for RPID are mostly $X(\gamma, n)X'$ reactions. The isotopes $X$ are identified by observing $\gamma$ rays characteristic of the reaction products $X'$.

(iv) In general there are several isotopes for a given element to be investigated. One may select one isotope $X$ and investigate the reaction product $X'$ with the appropriate life and the $\gamma$ decay schemes so as to get the best sensitivity for the element.

(v) Coincidence measurements of cascade $\gamma$ rays are very effective to get almost background-free measurements. A multidetector system is used to suppress Compton tails and to reduce background $\gamma$ rays and also to identify position/location of the impurity isotopes.

(vi) Laser electron photons, which are obtained from laser photons scattered off GeV electrons, are used to scan the sample with sub-mm position resolution.

*ejiri@rcnp.osaka-u.ac.jp

## III. REACTION RATES AND SENSITIVITIES OF RPID

RI production rate $R$ by the $(\gamma, n)$ reaction on $X$ is expressed as

$$R(X') = N_\gamma \cdot \sigma^e(\gamma, n) \cdot N(X), \tag{1}$$

where $N_\gamma$ is the photon intensity in the energy window $\Delta E$, $\sigma^e(\gamma, n)$ is the effective cross section for the photons in the energy window, and $N(X)$ is the total number of the isotopes $X$ per unit target area.

Then the number of $X'$ after photon irradiation for time $t$ is

$$N(X') = R(X') \cdot t_m \cdot \left[1 - \exp\left(-\frac{t}{t_m}\right)\right], \tag{2}$$

where $t_m$ is the mean life of $X'$. The $\gamma$ ray yield for the measurement time $t'$ is written as

$$Y(X') = \varepsilon(\gamma) \cdot \mathrm{Br}(\gamma) \cdot N(X') \cdot \left[1 - \exp\left(-\frac{t'}{t_m}\right)\right], \tag{3}$$

where $\varepsilon(\gamma)$ is the $\gamma$ detection efficiency and $\mathrm{Br}(\gamma)$ is the $\gamma$ branching ratio. The value for $\varepsilon(\gamma)$ is evaluated by using the EGS5 simulation code.

Hereafter, let us use practical units of $R(X')$ in units of $1/\mathrm{hour}$, $t$, $t'$, and $t_m$ in units of hours, $N_\gamma = n_\gamma \times 10^9/\sec$, $N(X) = (6 \times 10^{23}/A) \cdot n(X) \times 10^{-6}$ gr/cm², where $n(X)$ is the amount of $X$ in units of $\mu$ gr/cm² and $A$ is the mass number of $X$. The impurity in units of ppm is given as $i(X) = n(X)/n(T)$, where $n(T)$ is the weight of the sample in units of gr/cm².

The E1 giant resonant cross section is assumed to be given as $\sigma^e(\gamma, n) = 0.7\sigma(\gamma, n) = 2A$ mb, where $\sigma(\gamma, n) \approx 3A$ mb is the cross section at the resonance energy, and 1 b is $10^{-24}$ cm². Then one gets

$$N(X') = 4.3 \times 10^3 \cdot n_\gamma \cdot n(X) \cdot t_m \cdot \left[1 - \exp\left(-\frac{t}{t_m}\right)\right] \tag{4}$$

$$Y(X') = 4.3 \times 10^3 \cdot \varepsilon(\gamma) \cdot \mathrm{Br}(\gamma) \cdot n_\gamma \cdot n(X) \cdot t_m \cdot$$
$$\left[1 - \exp\left(-\frac{t}{t_m}\right)\right] \cdot \left[1 - \exp\left(-\frac{t'}{t_m}\right)\right]. \tag{5}$$

The sensitivity of RPID is defined as the minimum amount of the impurity isotopes, $n_m(X)$, to be detected by RPID. It is obtained from $Y(X') \geq \delta$, where $\delta$ is the number of minimum counts required to identify the peak with 68% C.L. It is given by $\delta \approx 2$ and $\delta \approx (B)^{1/2}$, with $B$ being the background (BG) counts at $E(\gamma)$, in cases of $B \leq 4$ and $B > 4$, respectively. The minimum impurity ratio $i_m(X)$ in weight is written as $i_m(X) = n_m(X)/n(T)$ ppm.

The minimum amount of impurity is given by $s \cdot n_m(X)$ with $s$ being the area of the photon beam spot on the sample. Laser electron photons scattered off GeV electrons are well defined in a small cone with $\theta \approx 1/\gamma_e$ with $\gamma_e$ being the $\gamma$ factor of the GeV electron beam. Then the beam spot is of the order of $s \approx 10^{-2}$ cm².

In fact, the sensitivity depends on the irradiation photon intensity, the lifetime of RI $X'$, the detector efficiency, and the BG. So, we evaluate the sensitivity for typical three cases of A, B, and C.

### A. Case A: Low-BG singles measurement

If the lifetime of $X'$ is much longer than those of other RIs produced by photonuclear reactions on other isotopes in the sample, and/or the $\gamma$ ray from $X'$ is higher in energy than other BG $\gamma$ rays, BGs at $E = E(\gamma)$ are assumed to be only the natural BGs. Using a low-BG Ge detectors with active shields like ELEGANT III, one gets $B \approx 0.02/\mathrm{hour}$ at $E(\gamma) = 0.5$–1 MeV [3]. Then, for a typical case of $t = t' = 2t_m$, $\varepsilon(\gamma) = 0.2$ with two Ge detectors and $\mathrm{Br}(\gamma) = 0.5$, $N(X')$, and $Y(X')$ are expressed as

$$N(X') = 3.7 \times 10^3 \cdot n_\gamma \cdot n(T) \cdot i(X) \cdot t_m, \tag{6}$$

$$Y(X') = 3.2 \times 10^2 \cdot n_\gamma \cdot n(T) \cdot i(X) \cdot t_m. \tag{7}$$

Then the sensitivity for $t_m = 10$ (hours) and $\delta = 2$ is obtained by requiring $Y(X') \geq 2$,

$$n(T) \cdot i_m(X) = 6 \times 10^{-4}/n_\gamma. \tag{8}$$

They are shown in Fig. 1. The minimum impurity for $n_\gamma \approx 1$ ($N_\gamma \approx 10^9/\sec$) is $N_m(X) = n(T) \cdot i_m(X) \approx 0.6$ ngr/cm² and $i_m(X) \approx 0.12$ ppb with $n(T) = 5$ gr/cm² sample. In case of $s \approx 0.01$ cm², the minimum amount of $X$ is $N_m(X) \cdot s = 6$ pgr.

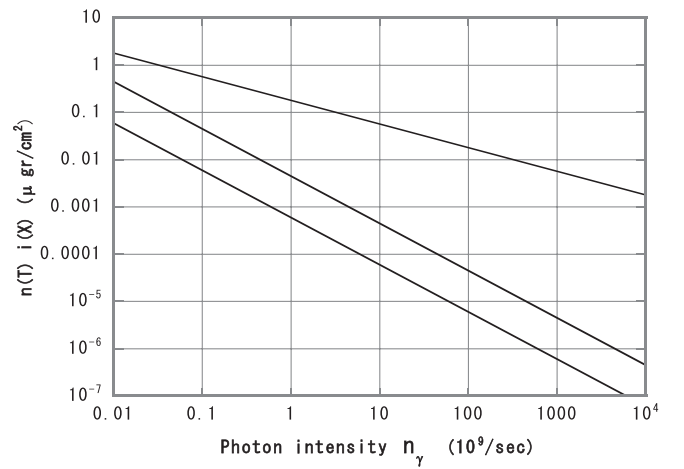

FIG. 1. Sensitivities of RPID in cases of A: low-BG singles measurements as given by Eq. (8); B: low-BG coincidence measurements as given by Eq. (9); and C: singles measurements with BGs from photonuclear reaction products of $n(T) = 5$ gr/cm² sample as given by Eq. (14), respectively.

### B. Case B: Low-BG coincidence measurements

In case of cascade $\gamma$ rays from $X'$, one can measure them in coincidence by means of two Ge detectors to reduce BG $\gamma$ rays. Then the BG will be almost 0 although the detection efficiency for 0.5–1 MeV $\gamma$ rays will be reduced to $\varepsilon(\gamma) \approx 0.02$. Then one gets for Br$(\gamma) = 0.5$, $t = t' = 2t_m$ and $t_m = 10$ hours,

$$n(T) \cdot i_m(X) = 6 \times 10^{-3}/n_\gamma. \tag{9}$$

Then the sensitivity for $N_\gamma = 10^9/\text{sec}$ photons is 6 n gr/cm$^2$ and 1.2 ppb in case of $n(T) = 5$ gr/cm$^2$. The total amount is 60 pgr in case of $s = 0.01$ cm$^2$. In case of RIs with a short mean life of $t_m \ll 10$ hours, one repeats many ($n \gg 10$) times of the photon irradiation and the $\gamma$ measurement for the short time intervals of $\Delta t = \Delta t' = t'/n \ll t_m$. For a typical case of $t_m = 1$ hour, $t = t' = 20$ hours, $n = 100$, $\Delta t = \Delta t' = 0.2$ hours, $\varepsilon(\gamma) = 0.02$ and Br$(\gamma) = 0.5$,

$$N(X') \approx 8.6 \times 10^4 \cdot n_\gamma \cdot n(T) \cdot i(X), \tag{10}$$

$$Y(X') \approx 4.3 \times 10^2 \cdot n_\gamma \cdot n(T) \cdot i(X). \tag{11}$$

In the present case of the coincidence measurements, the sensitivity is obtained by requiring $Y(X') \geq \delta = 2$,

$$n(T) \cdot i_m(X) \geq 4.5 \times 10^{-3}/n_\gamma. \tag{12}$$

Then one can measure impurity isotopes of the order of 5 ngr/cm$^2$ with $10^9$ photons/sec and 1 ppb in case of $n(T) = 5$ gr/cm$^2$. The total amount is 50 pgr with $s = 0.01$ cm$^2$.

### C. Case C: Singles measurements with BGs from other RIs

In the case that there are BG $\gamma$ rays from other RIs $S'$ produced by photonuclear reactions on other elements $S$ in the sample. They may be BGs in singles measurements if their energies are larger than $E(\gamma)$ and the lifetime of $S'$ is of the same order as $t_m$ of $X'$. Then the yield given in Eq. (11) is rewritten for the singles measurement as

$$Y(X') = 4.3 \times 10^3 \cdot n_\gamma \cdot n(T) \cdot i(X), \tag{13}$$

for $t = t' = 20$ hours, $\varepsilon(\gamma) = 0.2$, and Br$(\gamma) = 0.5$. BGs at $E = E(\gamma)$ are mainly Compton tails of $\gamma$ rays from the other RIs, and are suppressed much by using active shields. The BG yield for a typical case of Br$(\gamma) = 0.5$ and $i(S) = 10^5$ ppm is estimated as $Y(S') = 1.1 \times 10^5 \cdot n_\gamma \cdot n(T)$. Here we used the BG efficiency of $5 \times 10^{-5}$ for 3 MeV BG $\gamma$ rays with two Ge detectors. Then the sensitivity is obtained by requiring $Y(X') \geq [Y(S')]^{1/2}$ as

$$n(T) \cdot i_m(X) \geq 8 \times 10^{-2} \cdot [n(T)/n_\gamma]^{1/2}. \tag{14}$$

The sensitivity for $n(T) = 5$ gr/cm$^2$ is shown in Fig. 1. Even in this case of BG from RIs produced by RPIT on other isotopes $S$, it is possible to measure impurities of 180 ngr/cm$^2$ and 30 ppb, as shown in Fig. 1.

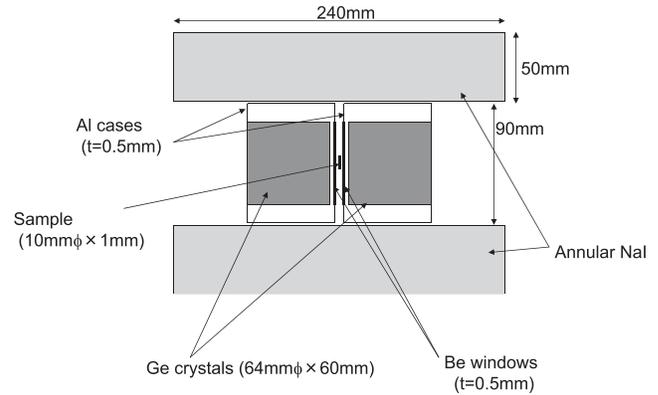

FIG. 2. A schematic view of a low-background $\gamma$ detection system with two Ge detectors and a cylindrical NaI active shield.

In fact the sensitivity is 1–3 orders of magnitude worse than case B with the coincidence measurement. Thus, coincidence measurements of $\gamma$ rays, x rays, and $\beta$ rays are better even if their efficiencies are small.

A possible high-sensitivity detector system may consist of two Ge detectors surrounded by a NaI active shield. The two Ge detectors are used for coincidence measurement of cascade $\gamma$ rays if BG in the singles mode is not very low, and otherwise for singles measurements to increase the efficiency. A schematic view of the detector arrangement with a thin sample is shown in Fig. 2. For a large sample of around 10 cm in width, it may be surrounded by many detectors to get sufficient coincidence efficiency. Such multidetector array is quite effective to keep the counting rate for the single detector below some detection limit.

## IV. SENSITIVITIES FOR MO-AU AND PB-U SAMPLES

### A. RPID for Mo-Au

In order to demonstrate the feasibility of RPID, a Mo-Au sample was irradiated by medium-energy photons, and $\gamma$ rays from $^{196}$Au RIs produced by RPIT on $^{197}$Au were measured by the ORTEC GMX45 Ge detector.

Medium-energy photons with $E_\gamma \approx 12$–16 MeV and $N_\gamma \approx 0.9 \times 10^6/\text{sec}$ were obtained from the 1.064 $\mu$m Nd-YVO$_4$ laser photons scattered off 0.95 GeV electrons at NewSUBARU. The Mo-Au with $i(^{197}\text{Au}) = 7.2 \times 10^4$ gm in weight were irradiated by the photons for $t = 8.9$ hours.

$\gamma$ rays from $^{196}$Au with $t_m = 214$ hours were measured by the Ge detector for $t' = 114$ hours after 20.6 hours from stopping the irradiation. The energy spectrum is shown in Fig. 3. The observed spectrum for $t' = 6$ hours run shows clearly 140.5 and 181 keV lines from $^{99}$Mo produced by $(\gamma, n)$ reactions on $^{100}$Mo and the 333 and 356 keV lines from $^{196}$Au produced by $(\gamma, n)$ reactions on $^{197}$Au.

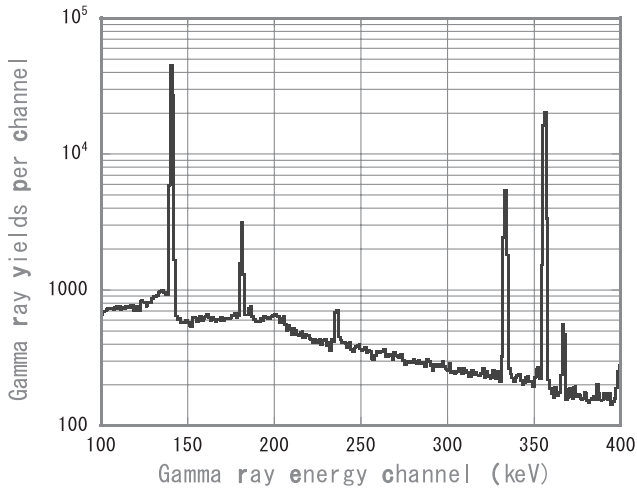

FIG. 3. Energy spectrum of $\gamma$ rays from natural Mo sample with 0.074 of Au in weight. The 140.5 and 181 keV lines are from $^{99}$Mo decays and 333 and 356 keV lines are from $^{196}$Au decays.

It is noted that the 356 keV peak is quite prominent, almost 3 orders of magnitude larger than the BG level even though the Ge detector is a conventional one. The total peak yield is $4.75 \times 10^5$, while the BG yield is $1.2 \times 10^3$ and the fluctuation is $[Y(BG)]^{1/2} \approx 35$. Noting this peak yield for the present Au ratio of $7.3 \times 10^4$ ppm, the peak yield would become 35 if the Au ratio would be 5.4 ppm. In other words, the sensitivity of the present conventional Ge detector is around 5 ppm for Au with the present $N_\gamma = 10^6/\text{sec}$. On the basis of the observed sensitivity with the present simple detector, let us evaluate the sensitivity for two cases.

(i) Using an active shield as in Fig. 3, the main BGs of the Compton tail of BG $\gamma$ rays from the sample Mo isotopes will be reduced by a factor 3, if active shields as given in Fig. 3 are employed to suppress the Compton tail. Then the sensitivity will be improved to 5.4 ppm/$\sqrt{3}$ = 3.1 ppm. If the photon intensity is increased by a factor $10^3$ to get $10^9/\text{sec}$, one may improve the sensitivity by a factor $\sqrt{(10^3)}$ to get 0.1 ppm, assuming the same signal to BG ratio. This is in accord with the sensitivity for C in Fig. 1.

(ii) Measuring the cascade 333 and 356 keV $\gamma$ rays in $^{196}$Au, the efficiency and the branching ratio of the cascade transition are $\varepsilon(\gamma_c) \approx 0.045$ and Br$(\gamma_c) \approx 0.21$. They are about 1/3 and 1/4 of the efficiency and the branching ratio for the singles measurement. Then the signals will be reduced by a factor 12. On the other hand, BGs are negligible and the minimum counts required for the peak identification are $\delta \approx 2$, which is a factor 17.5 smaller than the value of 35 for the singles measurement. Then the sensitivity will be $5.4 \times 12/17.5 = 3.7$ ppm. Increasing the photon intensity by a factor $10^3$, the signals will be increased by the same factor, while $\delta$ remains the same as $\delta = 2$. Then the sensitivity will be improved by a factor $10^3$ to be 3.7 ppb at $N_\gamma = 10^9/\text{sec}$. This is the same sensitivity as $B$ in Fig. 1.

## B. RPID for $^{238}$U

RPID may be used for nuclear fissile isotopes, long-lived fission products, and for difficult-to-measure isotopes ($\alpha$-emitter isotopes). $^{238}$U isotopes are measured by observing 208 keV $\gamma$ rays from $^{237}$U with $t_m = 234$ hours. They are produced via RPIT ($\gamma, n$) reactions. In the case of $t = 10$ hours, the number of $^{237}$U isotopes are obtained from Eq. (4) as $N(X') = 4.2 \times 10^4 \cdot n_\gamma \cdot n(T) \cdot i(X')$. The branching ratio of the 208 keV $\gamma$ rays is $B(\gamma) \approx 0.21$. Using two GMX Ge detectors, one gets $\varepsilon(\gamma) = 0.4$. Then, measuring for $t' = 10$ hours, the yield of the $\gamma$ rays is evaluated from Eq. (5) as $Y(X') = 150 \times n_\gamma \cdot n(T) \cdot i(X)$, i.e. $Y(X') = 150$ counts for $n(T) = 100$ gr, $i(X) = 10$ ppb and $N_\gamma = 10^9/\text{sec}$.

In case of U impurity in a 10 gr lead sample, major BGs are due to the Compton tail of the 279 keV $\gamma$ ray from $^{203}$Pb with $t_m = 75$ hours, which are produced by RPIT on $^{204}$Pb. Then one gets $N(^{203}\text{Pb}) = 4.0 \times 10^4 \cdot n_\gamma \cdot n(T) \cdot i(^{204}\text{Pb})$. The BG yield is $Y(^{203}\text{Pb}) = 2.4 \times 10^4 \cdot n_\gamma \cdot n(T)$ for the abundance ratio of $i(^{204}\text{Pb}) = 0.014 \times 10^6$ ppm, $\varepsilon(\gamma) = 4 \times 10^{-4}$, and Br$(\gamma) = 0.81$. The sensitivity is obtained by requiring $Y(X') \geq [Y(^{203}\text{Pb})]^{1/2}$. It is $n(T) \cdot i_m(X) \geq 1.0 \times [n(T)/n_\gamma]^{1/2}$, i.e., 100 ppb for $N_\gamma = 10^9/\text{sec}$ and $n(T) = 100$ gr.

The $^{238}$U impurity can be obtained by measuring the 609 keV $\gamma$ rays from the $^{214}$Bi, which is a chain isotope of $^{238}$U. The sensitivity is about the same as RPID, but the 609 keV BG line from $^{222}$Rn BG needs to be corrected for.

$^{235}$U isotopes can be detected by measuring fission product RIs by photofission reactions. Radioactive $^{99}$Mo isotopes with $t_m = 95$ hours are produced directly by the photofission and indirectly by $\gamma$ decays of the photofission products. Then the number of $^{99}$Mo isotopes is obtained by measuring 140.5, 181, and 739 keV $\gamma$ rays following $\beta$ decays of $^{99}$Mo. The photofission cross section on $^{238}$U is 1 order of magnitude smaller than that on $^{235}$U, and is 2 orders of magnitude smaller than the present photonuclear ($\gamma, n$) reaction on $^{238}$U [4]. $^{239}$Pu impurities, which are very rare in nature, may be measured by observing the 60 keV $\gamma$ rays from the photonuclear ($\gamma, 2n$) reaction on $^{239}$Pu. Identification of fissile isotopes of $^{235}$U/$^{239}$Pu via fission products are difficult [5]. Note that the present RPID emphasizes detection of very small impurities of the orders of ppb ($10^{-9}$) and ngr for science, and not for isotopes contained in heavy (10–30 cm) shields.

## C. RPID for fission products: $^{90}$Sr

Many kinds of stable and radioactive nuclei are produced by nuclear fissions. Some of them are $\alpha$ and/or $\beta$ emitters. and are difficult to be measured by $\gamma$ detectors. They are measured by transmuting them to $\gamma$-emitter nuclei. Since fission products in fissile nuclei are of the order of 100–1000 ppm, one can measure them by using conventional photon sources with $N_\gamma \approx 10^6$/sec.

$^{90}$Sr isotopes, which are $\beta$-emitter nuclei, are transmuted to $^{89}$Rb isotopes with $t_m = 22$ min. via $(\gamma, p)$ reactions. Then $^{89}$Rb isotopes are measured by observing the 1032 keV $\gamma$ rays. Using Eq. (7), the $\gamma$ ray yield for $t = t' = 2t_m$ is

$$Y(X') = 4 \times n_\gamma \cdot i(X) \cdot n(T),  \qquad (15)$$

where $i(X) \cdot n(T)$ is the $^{90}$Sr isotopes in units of $\mu$gr in the sample, and the $(\gamma, p)$ cross section is assumed to be about 0.05 $\sigma(\gamma, n)$ and $\epsilon(\gamma) = 0.12$, Br$(\gamma) = 0.58$. Then using $n_\gamma = 10^{-3}$ in units of $10^9$ photons/sec and $n(T) = 1$ kg sample with 100 ppm $^{90}$Sr, the yield is $4 \times 10^2$. Therefore it is easy to measure $^{90}$Sr with the concentration of the order of 100 ppm.

## V. REMARKS AND DISCUSSIONS

The present RPID with a realistic low-background Ge detector system is a high-sensitivity nondestructive assay of the order of ppm-ppb by means of $N_\gamma = 10^9$/sec, and even those of ppt level by using intense photons of the orders of $N_\gamma = 10^{12-15}$/sec.

Laser electron photons with the very small beam spot make it possible to investigate impurities in 10–20 mg samples, and to scan and identify them with sub-mm position resolution.

In fact the sensitivity depends much on background $\gamma$ rays from RIs produced by photonuclear reactions on the sample. If they are appreciable, coincidence measurement is crucial for high-sensitivity assay.

The present RPID uses medium-energy photons in a wideband of $\Delta \approx 5$ MeV, i.e., effectively a half of incident photons. The integrated cross section over the resonance is as large as $\sigma(\Gamma) \sim 2$ b MeV for nuclei with $A \sim 200$, and photons used for RPID are as intense as $n_p = 10^8 - 10^{12}$/MeV/sec from the photon sources of $10^9 - 10^{13}$/sec. Then the product is $\sigma(\Gamma)n_p = 2 \times 10^8 \sim 2 \times 10^{12}$ b/sec.

The nuclear resonance fluorescence (NRF) is also used to study nuclear isotopes [5–7]. The NRF peak cross section is very large, but the natural width $(\Gamma)$ is very small. Then the integrated cross section over the resonance is of the order of $\sigma(\Gamma) = 10^{-2}$ b keV $(10^{-5}$ b MeV). Doppler broadening makes the width broader, but the integrated cross section remains the same. This is useful for large amount (more than $\mu$gr) isotopes with large $\Gamma$ levels as fissile isotopes. By using intense photons of $10^{10}$/keV/sec from initial $10^{13}$/sec photons [5], the product is $\sigma(\Gamma)n_p = 10^8$ b/sec. Intense photons of $10^6$/eV/sec in the narrow band of 0.1% is expected at ELI-NP and MEGa-Ray [8]. RPID and NRF measures nuclear $\gamma$ rays characteristic of photonuclear reaction products from the isotopes to be studied and of the isotopes themselves, respectively. Thus both select specific isotopes and the $\gamma$ yields depend on the isotopes. Attenuation of RPID 15 MeV photons are mainly due to the pair creation, while that of NRF 2 MeV photons is mainly due to Compton effects. They are similar for medium and heavy nuclei. NRF measures prompt $\gamma$ rays, while RPID does delayed $\gamma$ rays from residual radioactive isotopes. Then RPID can be free from prompt BG $\gamma$ rays and delayed ones from short-lived RIs. In any way RPID and NRF are complimentary to each other.

Photon sources available at present are HI$\gamma$S with FEL [9], NewSUBARU with a Nd-YVO$_4$ laser [1], and others. The HI$\gamma$S photon intensity is of the order of $N_\gamma \approx 10^9$/sec with 50 mA 0.475 GeV electrons and 1.6 eV FEL laser photons [9]. It will be of the order of $N_\gamma \approx 10^{12}$/sec with 500 mA and 1.2 GeV electrons.

New generation synchrotrons at 3 GeV and intense CO$_2$ laser will provide $N_\gamma \approx 10^{10}$/sec. New generation photon sources with $N_\gamma \approx 10^{13-15}$/sec are planned [4,10–14]. These are very promising for high-sensitivity ppt level studies of nuclear isotopes. RPID using a narrow beam obtained by the Compton scattering off low-emittance electron beam has good position resolution of the order of 1 mm. The large index of reflection [15] of silicon for $\gamma$ rays is of great interest for RPID with better position resolution by focusing photons with the refractive optics technique.

Bremsstrahlung photons induced by medium-energy electrons from conventional betatrons and/or linear accelerators can be also used for RPID with ppm sensitivity. The efficiency is low since only a small fraction of the bremsstrahlung photons is used to excite E1 giant resonance, and the beam spot is not very small. Intense linear accelerators are powerful. Fission products and other isotopes with 100–10 ppm level are easily measured by using simple $\gamma$ sources with $N_\gamma \approx 10^6$ sec.

Recently medium-energy $[E(\gamma) \approx 10–50$ MeV] photons are available by using high-power lasers [16]. They are produced by medium-energy electrons following interaction of peta $W$—sub-peta $W$ laser pulses with heavy metal target. It should be emphasized that RPIT does not require high energy resolution (narrow band), well collimated (narrow emittance) photon beams in contrast to nuclear physics experiments. Medium-energy photons from high-power lasers are of great interest for RPID as well as RPIT [1].

## ACKNOWLEDGMENTS

The authors thank Dr. M. Fujiwara and Dr. A. Titov for valuable discussions.